\documentclass[chicago]{emulateapj}

\usepackage{txfonts}
\usepackage{longtable}
\usepackage{rotating}
\usepackage{natbib}
\usepackage{graphicx}
\usepackage{graphics}
\usepackage{psfrag}
\usepackage{amssymb}
\usepackage{mathrsfs}
\bibpunct{(}{)}{;}{a}{}{,}
\usepackage{color}
\usepackage{ctable}
\usepackage{tablefootnote}
\usepackage{natbib,twoopt}
\usepackage{threeparttable}

\def\Teff{$T_{\mathrm{eff}}\,$}

\def\Pt{\ensuremath{\rm P_{\mathrm{turb}}}}
\def\Msun{\ensuremath{\rm \,M_\odot\,}}
\def\logtau{\ensuremath{\rm \,log(\tau)\,}}
\def\Ll{{\mathscr L}/{\mathscr L_\odot}}
\def\logLl{\rm log(L/L{\odot}}

\shorttitle{Turbulent pressure in massive star envelopes}
\shortauthors{Grassitelli et al.}

\begin{document}


\title{Observational consequences of turbulent pressure in the envelopes of massive stars}


\author{L.Grassitelli $^1$}
\email{luca@astro.uni-bonn.de}
\author{L.Fossati $^1$}
\author{S.Sim{\'o}n-D{\'{\i}}az $^{2,3}$}
\author{N.Langer $^1$}
\author{N.Castro $^1$}
\author{D.Sanyal $^1$}
\affil{$^1$Argelander-Institut f\"ur Astronomie der Universit\"at Bonn, Auf dem H\"ugel 71, 53121, Bonn, Germany}
\affil{$^2$Instituto de Astrof\'{i}sica de Canarias, 38200 La Laguna, Tenerife, Spain}
\affil{$^3$Departamento de Astrofisica, Universidad de La Laguna, 38205 La Laguna, Tenerife, Spain}

%
\begin{abstract}
The major mass fraction of the envelope of hot luminous stars is radiatively stable.
However, the partial ionisation of hydrogen, helium and iron gives rise to extended sub-surface
convection zones in all of them. In this work, we investigate the effect of the pressure
induced by the turbulent motion in these zones based on the mixing length theory, 
and search for observable consequences.
We find that the turbulent pressure fraction can amount up to $\sim 5\%$ in OB supergiants,
and to $\sim30\%$ in cooler supergiants. The resulting structural changes are, however, not significantly affecting the evolutionary tracks compared to previous calculations. Instead, a comparison of macroturbulent velocities derived
from high quality spectra of OB stars with the turbulent pressure fraction obtained in corresponding
stellar models reveals a strong correlation of these two quantities. We discuss a possible physical
connection, and conclude that turbulent pressure fluctuations may drive high-order oscillations, 
which --- as conjectured earlier --- manifest themselves as macroturbulence in the photospheres of 
hot luminous stars.   

\end{abstract}


\section{Introduction}

Massive stars are of key importance for the enrichment of the interstellar medium \citep{2006Kobayashi}, for regulating star formation \citep{2004MacLow}, and as progenitors of supernovae and gamma ray bursts \citep{2012Langer}. However, various physical processes acting in massive stars are as yet not well understood, therefore preventing the current stellar models to self-consistently explain several general observational properties, such as the mass-discrepancy \citep{1992Herrero,2015Markova}, the presence of the Humphreys-Davidson (HD) limit \citep{1979HD}, and the position of the terminal-age-main-sequence in the HR diagram \citep[e.g.,][]{2010Vink,2014Castro}. Radiation pressure dominated layers, envelope inflation, clumpy stellar winds, and dynamical instabilities are features which commonly occur in these objects, but which are not yet thoroughly investigated.
Some of these phenomena have been connected to the presence of turbulent motions in the outer 
layers \citep[e.g.,][]{1984deJager,2009Cantiello}. Despite its simplicity, the mixing-length theory \citep[MLT,][]{1958Vitense} for convection has
been very successful in describing the main features of non-adiabatic turbulent stellar envelopes \citep[e.g.,][]{2014Trampedach}.
However, the MLT is expected to have shortcomings when the convective velocities approach the speed of sound, as it is expected in the envelopes of very luminous stars \citep{2015Sanyal}. In particular, it has been argued by \citet[]{1984deJager} and \citet{2009Maeder} that turbulent pressure may become important in this situation.  It is the aim of the present paper to discuss the role of turbulent pressure in the envelopes of massive stars, investigate its effects on the stellar structure, and focus on its potential observational signatures.

\section{Method}

We use the Lagrangian one-dimensional hydrodynamic stellar evolution code BEC \citep{2000Heger,2005Petrovic,2011Brott} for 
computing massive star models. It treats convection following the MLT 
with a mixing-length parameter of $\rm\alpha=1.5$ \citep{2011Brott}. 
The opacity of stellar matter is interpolated from the OPAL opacity tables \citep{1996Iglesias}, and mass-loss by stellar wind 
is adopted following the prescription of \citet{2001Vink}.       

We modified the BEC code in order to include the turbulent pressure term in the stellar envelopes, in particular in the convective zones 
associated with the iron opacity peak (FeCZ), and with the partial ionisation of helium (HeCZ), and hydrogen (HCZ). 
We neglect the turbulent pressure in the convective core, where it is typically more than 7 orders of magnitude smaller than the 
ideal gas pressure. We included the turbulent pressure $\Pt$ and the turbulent energy density $\rm e_{turb}$ in the momentum, energy transport and
energy conservation equations, respectively, following \citep{1991Canuto,1997Jiang,2003Stothers,2009Maeder,2014Trampedach}:
\begin{equation}
P_{turb}=\zeta\rho v_c^2\quad,\quad e_{turb}=\frac{3}{2}\frac{P_{turb}}{\rho},
\end{equation}
where $\rho$ is the local mass density, $\zeta$ is a parameter chosen to be $\zeta=1/3$ for isotropic turbulence \citep{2003Stothers,2009Maeder}, 
and $v_c$ is the local convective velocity. As the convective velocities approach the local sound speed, we limit the convective velocities
to this value, i.e. 
\begin{equation}\label{convel}
v_c\leq c_s\quad\quad,\quad c_s^2= \frac{k_B T}{\mu m_H}=\frac{P_{gas}}{\rho},
\end{equation}
where $T$ is the local temperature, $k_b$ is the Boltzmann constant, $\mu$ is the mean molecular weight, and $m_H$ is the proton mass.
Here, we use the isothermal sound speed, as the layers under consideration are characterised by
a small ratio of the local thermal-to-dynamical time scale ($\tau_{th}/\tau_{dyn}<1$).
As a consequence of Eq.\ref{convel}, the turbulent pressure in our models can not exceed a value of one third of the local gas pressure.
\newpage
\section{Results}

\begin{figure}[!]
\resizebox{1\hsize}{!}{\includegraphics{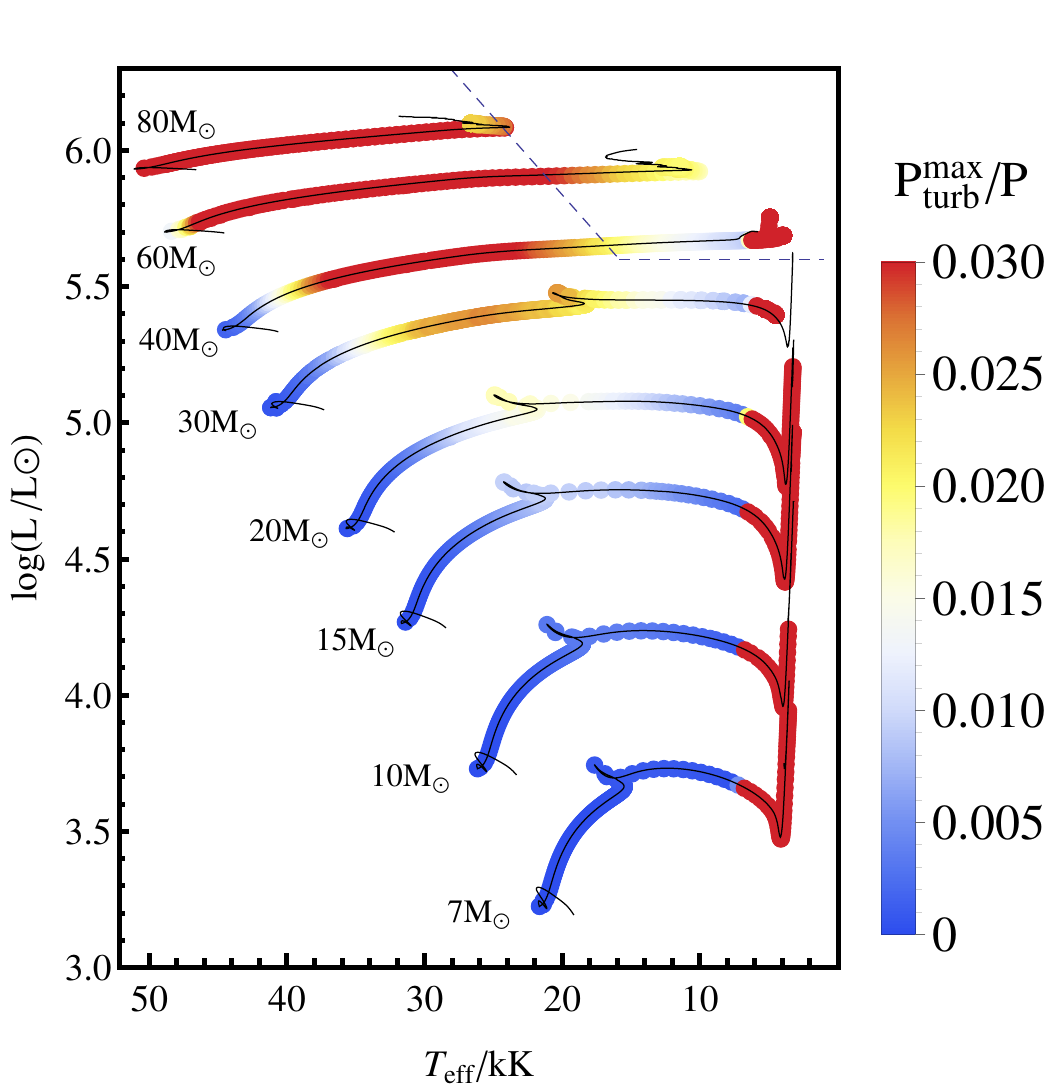}}\caption{Stellar evolution tracks calculated including turbulent pressure and turbulent energy density,
indicated by colored dots. 
Superposed are the tracks by \citet{2011Brott} (black lines). The colour indicates the maximum fraction of turbulent pressure occurring within each stellar model (see color bar to the right), where fractions above 3\% are indicated in red. Stellar models at \Teff$<10\,$kK coloured in red show contributions from $\Pt$ up to $\approx33\%$ and hot luminous OB stars coloured red up to $\approx5\%$. The stellar masses next to the tracks indicate the initial mass of the models. 
The dashed line indicates the position of the HD-limit \citep{1979HD}.}
\label{fig:HRPturb}
\end{figure}

We computed a set of stellar models and stellar evolutionary tracks for non-rotating stars with Galactic metallicity
in the initial mass range 7--80\Msun, including the physics described in Sect.~2. All input and physics parameters
were chosen as in \citet{2011Brott}.

\subsection{Evolutionary Tracks}

The evolutionary tracks of our newly computed models are plotted in Fig.~\ref{fig:HRPturb} and superposed to those obtained by \citet{2011Brott}.
Throughout most of the evolution, the effect of turbulent pressure and energy on the luminosity and the surface temperature
of the models is small, such that our new evolutionary tracks nearly coincide with those from \citet{2011Brott}. 
According to the colour scheme in Fig.~\ref{fig:HRPturb}, which indicates the maximum fraction of turbulent pressure inside the
stellar models, the HR-diagram can be roughly divided into three areas: the hot (\Teff$>10^4$\,K) and luminous 
($\logLl)>4.5$) stars, the cool stars (\Teff$<10^4$\,K) of any luminosity, and the hot stars with
$\logLl)<4.5$.

In the latter regime, the turbulent pressure does not exceed a fraction of the total pressure of a few tenths of a percent anywhere inside
the stellar models. On one hand, the iron opacity peak in these models is located deep inside the envelope in nearly adiabatic layers, 
which leads to relatively low convection velocities. On the other hand, the models are
too hot to contain a hydrogen ionisation zone. Finally, they do show a HeII partial ionisation zone close to the surface, which
is, however, not vigorous enough to play a significant role \citep{2009Cantiello}.      

In the cool supergiant region, the turbulent pressure fraction inside the star can be very significant. 
The maximum value is typically 25--30\%, while the highest possible value, 33\%, is reached for the most luminous models.
The computed transonic convective velocities, of the order of 20--30\,km/s, arise in the HCZ very close to the surface, 
where high opacities induce high local Eddington factors, density inversions, 
and high degrees of superadiabaticity \citep{2015Sanyal}. Such a high turbulent pressure fraction is possible here
since ideal gas pressure is by far the dominant contribution to the total pressure in the envelopes of the cool supergiants. Even though a significant fraction of the total pressure arises from the turbulent motion, the evolutionary tracks are only shifted by few tens of 
Kelvin towards lower temperatures, compared to the tracks without turbulent pressure, which corresponds to a radius increase
by a few percent.
This is due to the fact that the region which contains high turbulent pressure contains very little mass (see below). 

In the hot and very luminous stars, the turbulent pressure can account for up to $\sim$5\% of the total pressure. The evolutionary tracks are 
not significantly different from those of \citet{2011Brott}, showing displacements of the order of hundreds of Kelvin towards lower 
effective temperatures. Turbulent pressure becomes more important as the stars expand during their main sequence evolution,
as the convective velocities increase.
It achieves a maximum in the O supergiant regime, and then decreases in the B supergiant regime as the 
iron opacity peak moves deeper inside, but turbulent pressure remains significant for
surface temperature above $\sim$10\,000\,K.
The maximum turbulent pressure in these models occurs within the FeCZ, where high local Eddington-factors are achieved, giving rise to envelope inflation and density inversions \citep{2015Sanyal}.
It does not reach as high fractions as in the cool supergiant because of the predominance of radiation pressure in the hot star envelopes.

\subsection{The structure of the outer layers}

\begin{figure}
\vspace{-0.0cm}
\resizebox{1.0\hsize}{!}{\includegraphics{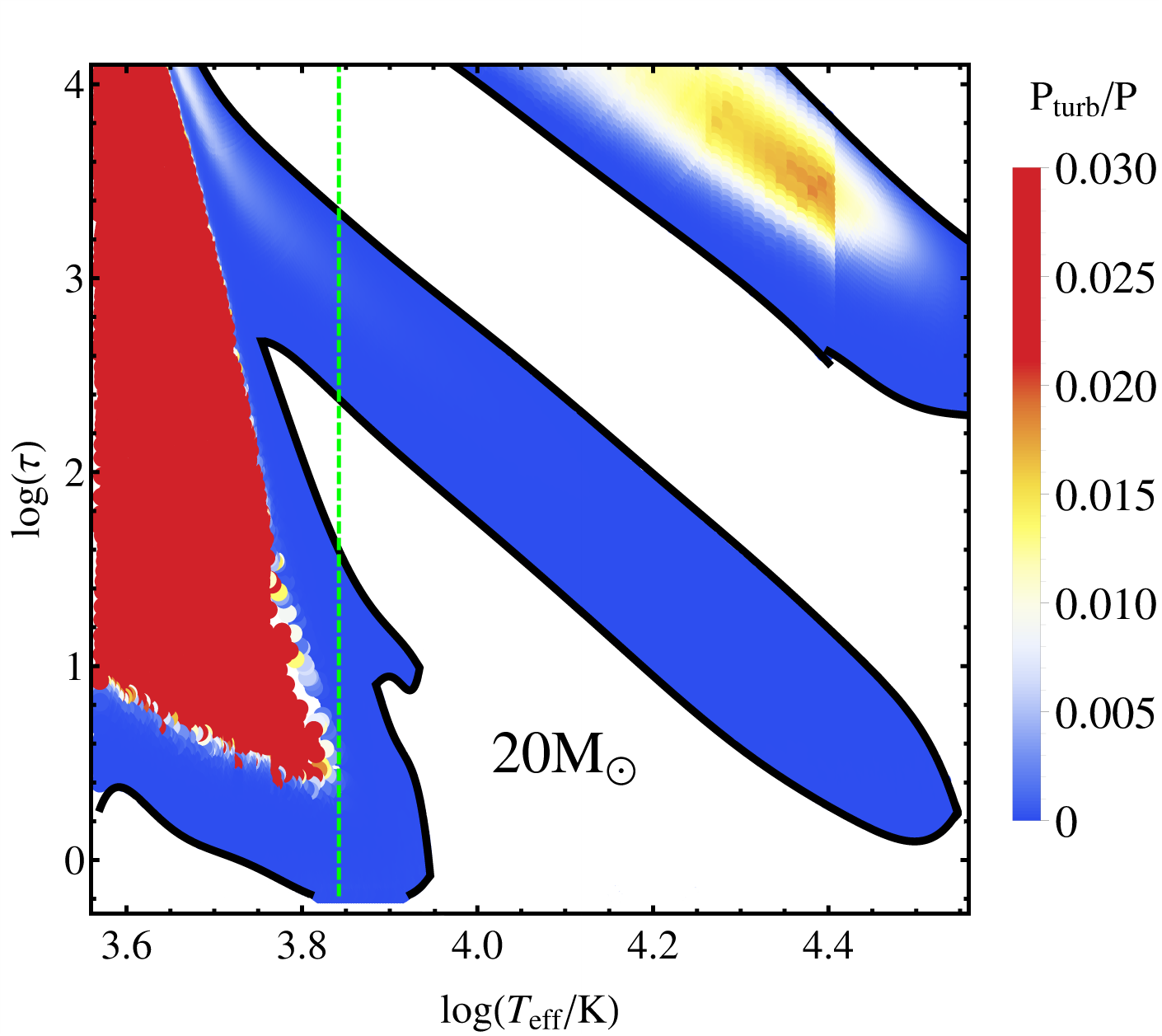}}
\newline
\resizebox{1.0\hsize}{!}{\includegraphics{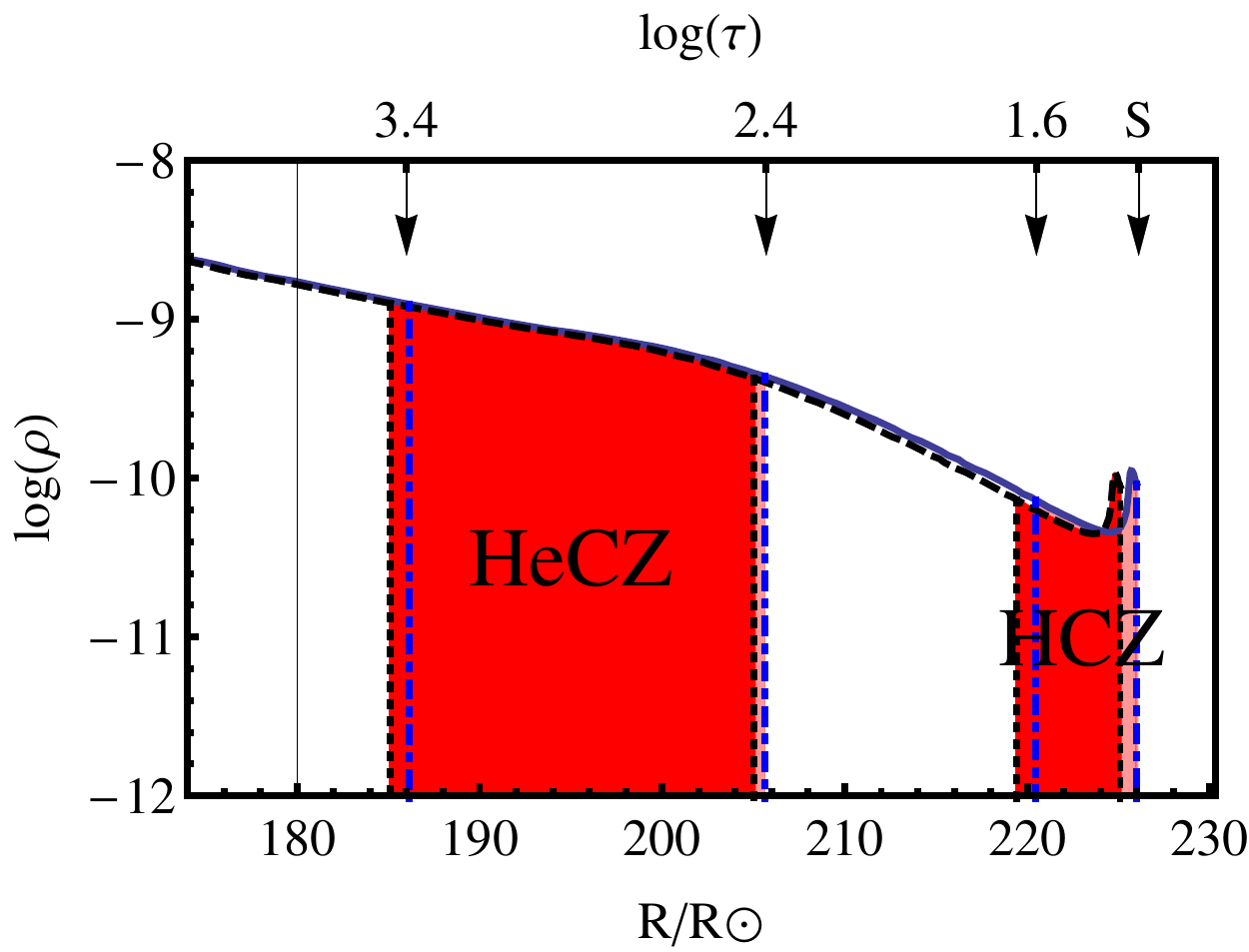}}
\caption{Top: ratio of turbulent-to-total pressure (color coded) as a function of effective temperature and optical depth
throughout the evolution of the 20\Msun model. The FeCZ is visible in the upper-right corner, the HeCZ lies across the diagram, and the 
HCZ appears for \Teff  below $\sim$10\,000\,K. The dashed line indicates the selected density profile shown in the bottom panel.  
\newline
Bottom: Comparison of the density (in g\,$\rm cm^{-3}$) as a function of the radial coordinate R in the outer layers 
of our 20\Msun star at \Teff$\simeq7000\,$K (blue solid line) with that in an identical model where the turbulent pressure term was switched off (black dashed line).
The shaded dark-red regions confined within the dotted vertical lines denote the HeCZ and HCZ in the model calculated without \Pt, 
while the light-red regions correspond to the same convection zones in the model calculated with \Pt. The numbers associated to the arrows on the top of the figure indicate the optical depth \logtau of the borders of the convective zones in the case with \Pt included (with S indicating the surface). 
}
\label{fig:20}
\end{figure}

\begin{figure}
\resizebox{1.00\hsize}{!}{\includegraphics{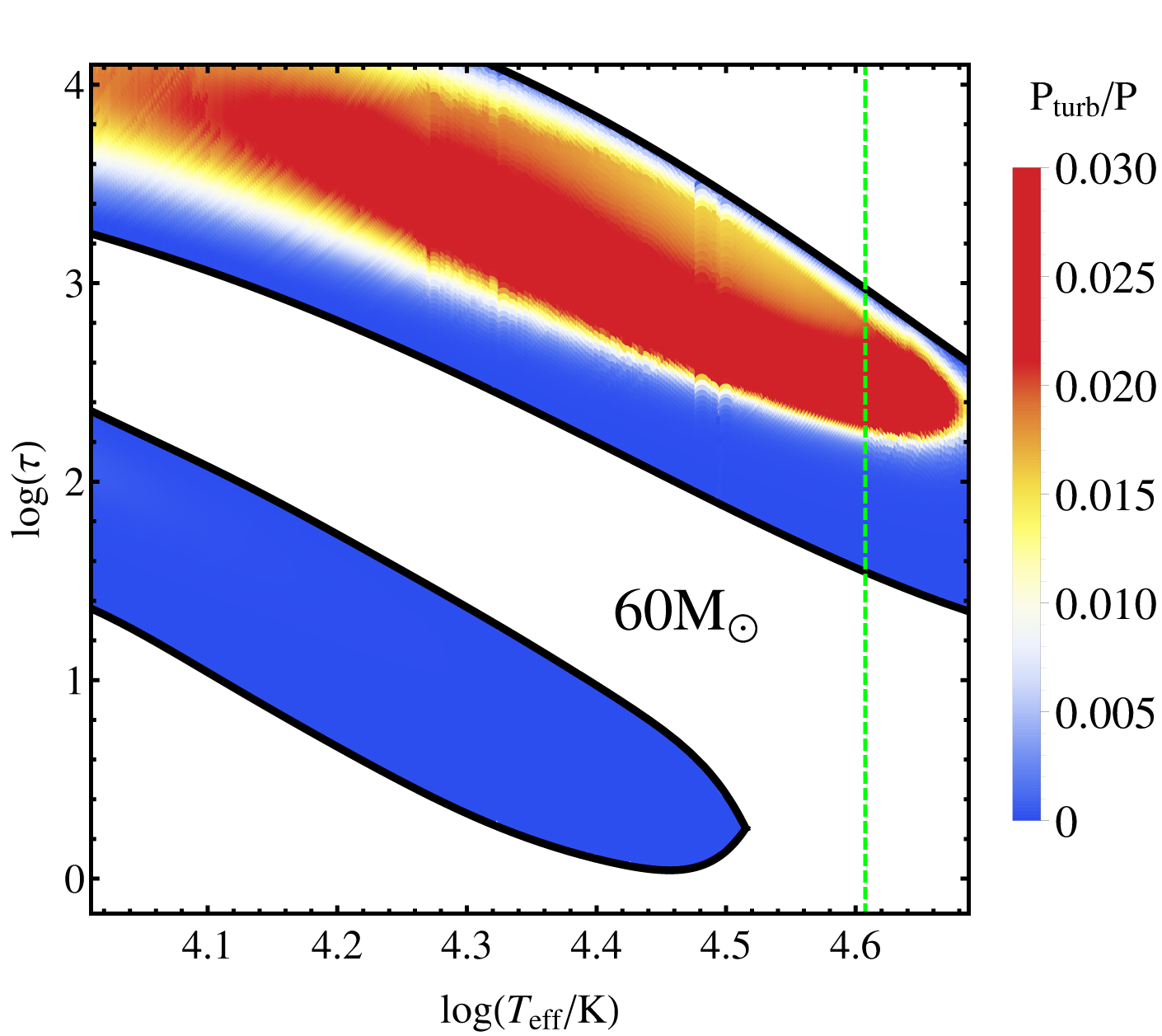}}
\newline
\resizebox{1.0\hsize}{!}{\includegraphics{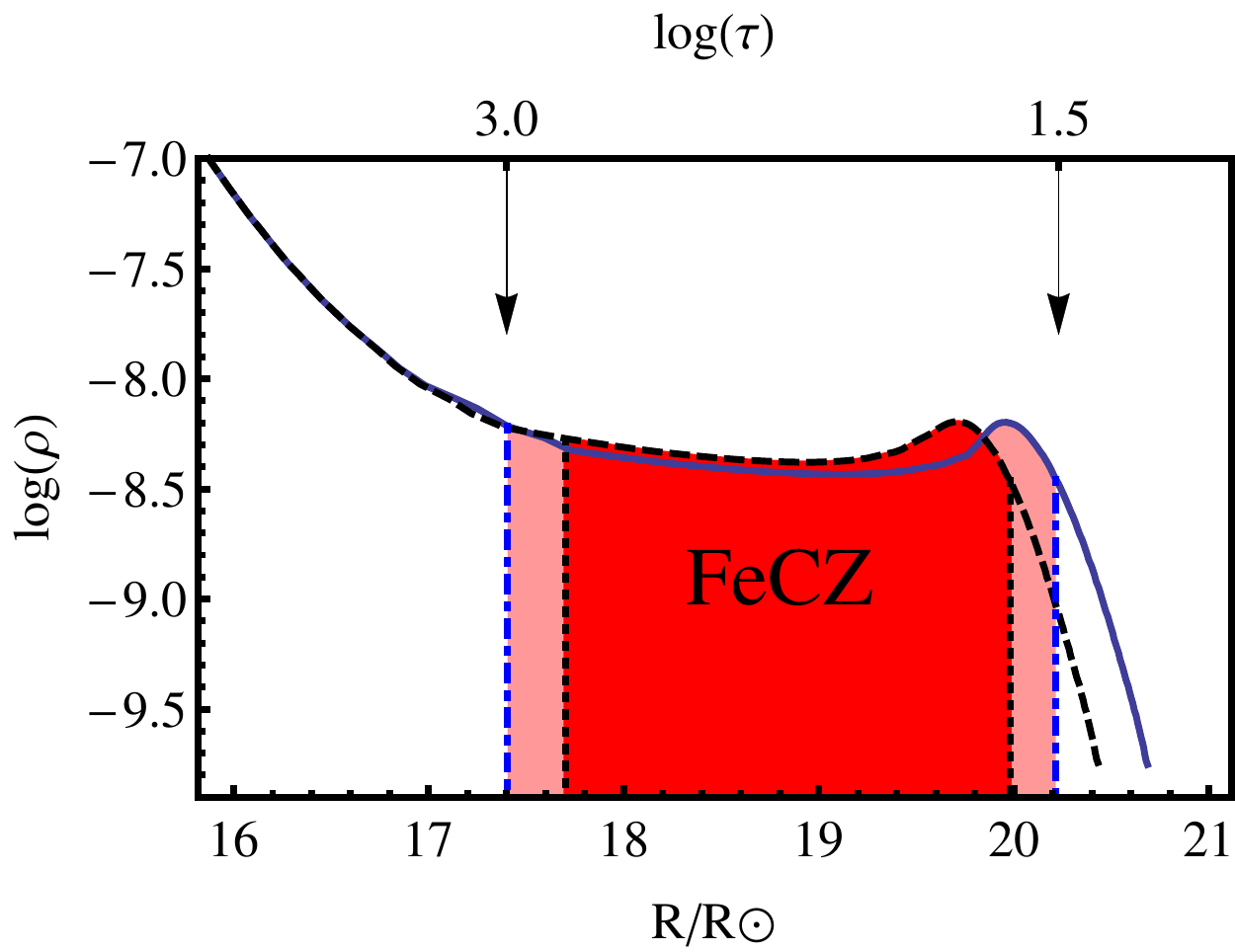}}
\caption{Same as Fig.~\ref{fig:20}, but for the 60\Msun model. Top: ratio of turbulent-to-total pressure (color coded) as a function of effective temperature and optical depth
during the evolution of the 60\Msun model, showing 
the FeCZ and HeCZ are at higher and lower optical depths, respectively.
\newline Bottom: density profile
and extent of the FeCZ at \Teff$\approx40\,000\,$K. 
}
\label{fig:60}
\end{figure}

Following the results shown in Sec~3.1, we investigate the structure of the 20\Msun and 60\Msun models
in more detail. Figs.~\ref{fig:20} and~\ref{fig:60} show the relative fraction of the turbulent pressure
as function of the optical depth in the stellar envelopes of both models throughout the evolution.

In the 20\Msun model (Fig.~\ref{fig:20}), we find that for \Teff$\geq10^4\,$K the turbulent pressure fraction
has its maximum in the FeCZ, located at an optical depth of \logtau$\approx2.5-4$. The turbulent pressure 
fraction is about 1.8\% at \Teff$\simeq25\,000\,$K, and the FeCZ moves deeper 
inside the star as the stellar model expands during its evolution. 
Once the stellar model reaches effective temperatures well below 10\,000\,K, the HCZ arises. 
This convective region, within which the maximum turbulent pressure fraction arrives at about 25--30\%, reaches the stellar
surface in stellar models with \Teff$<8000\,$K,
and covers an extended range of optical depths. 
The bottom panel of Fig.~\ref{fig:20} compares the density profiles and the radial extent of the convection zones 
for a model with  \Teff$\simeq7000\,$K and turbulent pressure included to the same model where turbulent pressure was disregarded. 

In the envelope of the 60\Msun model, we find convective zones associated 
to the partial ionisation of iron and helium, while the model does not become cool enough to show hydrogen recombination. 

As shown in Fig.~\ref{fig:60}, the FeCZ is located at an optical depth of \logtau$>1.5$, 
with $\rm\Pt/P$ rising to $\sim$5\% for \Teff$\simeq40\,000\,$K. As the star evolves the convective region moves deeper inside the star. 
Figure~\ref{fig:60} shows that due to the turbulent pressure, the star increases its radius by few percent, 
which leads to a slightly reduced density and an increase of the radial extent of the iron convection zone. 

\section{Comparison with observations}

In particular the iron opacity peak at T$\,\approx\,$20\,0000\,K is known to induce a variety of
dynamical phenomena at the stellar surface,
e.g. pulsations in $\beta$-Cephei and slowly pulsating B stars \citep[SPB, ][]{2007Miglio,1999Pamyatnykh}, and stochastically excited 
travelling waves generated by turbulent motions in the FeCZ  \citep{1990Goldreich,2009Cantiello,2010Belkacem,2010Samadi,2014Mathis} 
leading to a subsonic small scale velocity field at the surface that has been proposed to be the physical origin of the so called ``microturbulence''.

The spectra of luminous O-B stars is known to be also affected by the so called macroturbulent broadening, an extra line-broadening usually ad-hoc associated with large scale (compared to the line forming region) motions at the surface (see \citet{2014Markova}, \citet{2014SimonDiaz}, and references therein). Similarly to the case of microturbulence, convection might play a significant role in the origin of macroturbulence as well. 
This view is supported by the work of \citet{2013Sundqvist}, who showed that macroturbulence is suppressed in strongly magnetic massive stars, where the magnetic field is expected to at least partially inhibit convection. 
A similar effect was found for intermediate mass magnetic chemically peculiar stars \citep{1997Ryabchikova} and for spots in late-type stars (Strassmeier 2009). In this context, vigorous envelope convection in the temperature range of the hydrogen recombination (Fig.~\ref{fig:20}) may be responsible for the non-thermal (macroturbulent) broadening observed in red supergiants \citep{2007Collet,2008Carney}. 

With this motivation, we pursue the hypothesis that the relative strength of turbulent pressure in the sub-surface convective zones
is related to the appearance and strength of macroturbulence at the stellar surface. 
We investigate the case of the luminous O-B stars, where the turbulent pressure constitutes 
up to $\approx$5\% of the total pressure in the FeCZ. 

We make use of the results from the quantitative spectroscopic analysis of the rich sample of spectra compiled by the IACOB project \citep{2011IACOB,2015IACOB}. In particular we benefit from the derived values of surface temperature (\Teff), gravity (log(g)), projected rotational velocity (vsin(i)) and macroturbulent velocity ($v_{macro}$) for a sample of $\sim$~300 Galactic O-B stars used in \citet{2015SimonDiaz}\footnote{The results of the analysis of a much larger dataset, along with a thorough empirical description of the behaviour of macroturbulent broadening in the whole O-B star domain will be presented in Sim{\'o}n-D{\'{\i}}az et al. (in prep).}. 

\begin{figure*}
\resizebox{1\hsize}{!}{\includegraphics{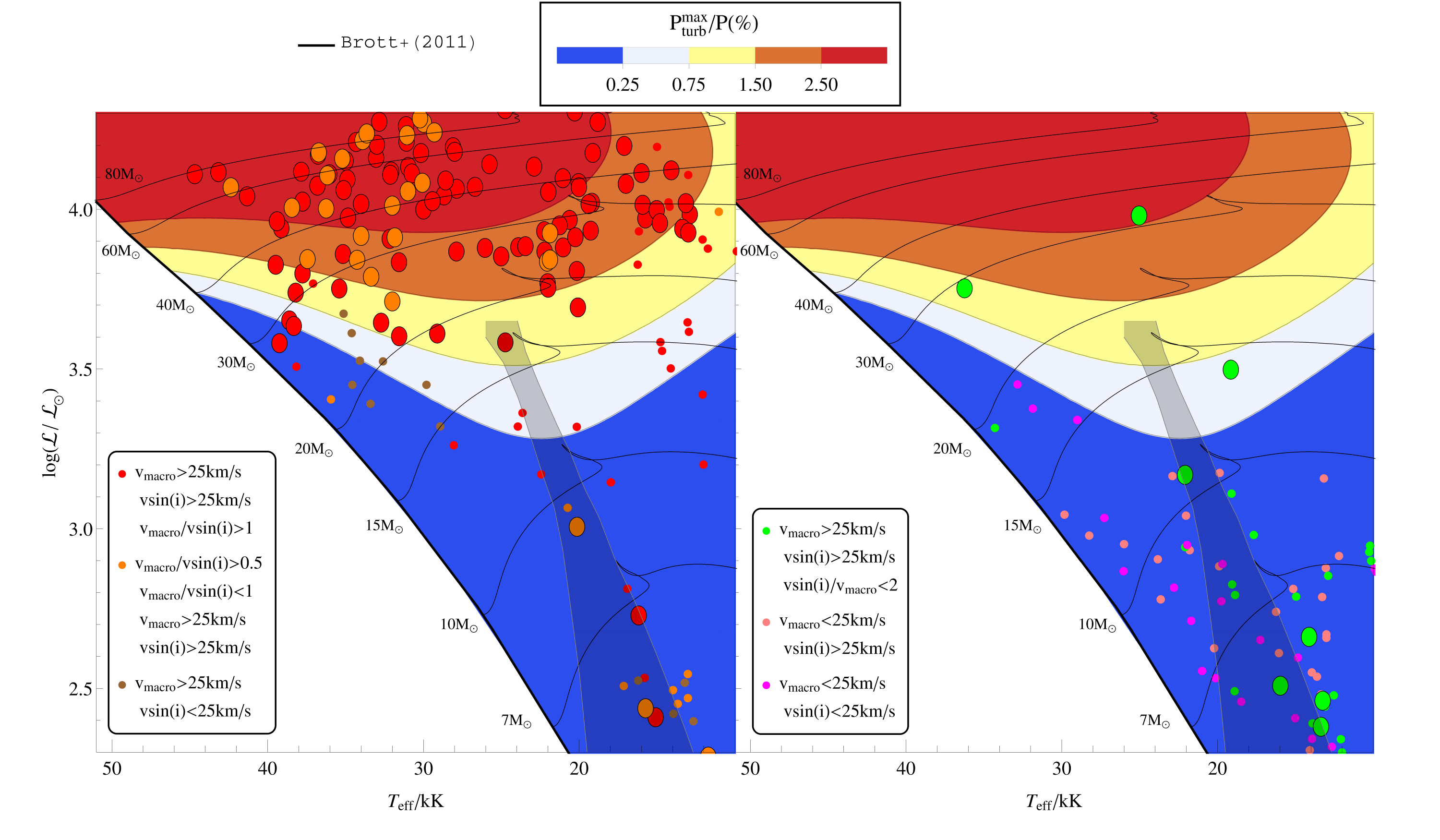}}
\caption{Spectroscopic HRD with coloured regions representing the maximum fraction of turbulent pressure as derived from a best-fit of the tracks in Fig.~\ref{fig:HRPturb}. Circles represent the observed O-B stars, colour coded according to their spectral line shape as in \citet{2015SimonDiaz}, i.e. following the ratio between the projected rotational velocity $v\,sin\, i$ and the macroturbulence velocity $v_{macro}$. In the left panel stars with line profiles showing a clear contribution from macroturbulence are located, while the right panel includes stars with line profile showing a dominant rotational broadening and a less clear contribution from macroturbulence. The bigger circles bordered in black indicate stars showing $v_{macro}$  higher than 50 km/s.  The gray bands indicate the $\kappa-$mechanism instability strip for SPB stars \citep{2007Miglio}.  }
\label{fig:sHRmacro}
\end{figure*}

In Fig.~\ref{fig:sHRmacro} we compare the observationally derived macroturbulent velocities to the maximum fraction of turbulent pressure in our models in the spectroscopic HRD \citep[sHR,][]{2014Langer}. Stars presenting a clear signature of macroturbulence in their 
line profiles (i.e., a $v_{macro}$ larger than 50\,km/s) are marked by a bigger black circle in Fig.~\ref{fig:sHRmacro}.
Interestingly, those stars are located mainly in regions of the sHR where the turbulent pressure yields the highest contribution to the total pressure 
in the FeCZ. At places where the models predict a small contribution from the turbulent pressure only very few stars show an unambiguously  
high macroturbulent velocity. We discuss these exceptions at low ${\mathscr L}$ in the next section.

\begin{figure}
\resizebox{\hsize}{!}{\includegraphics{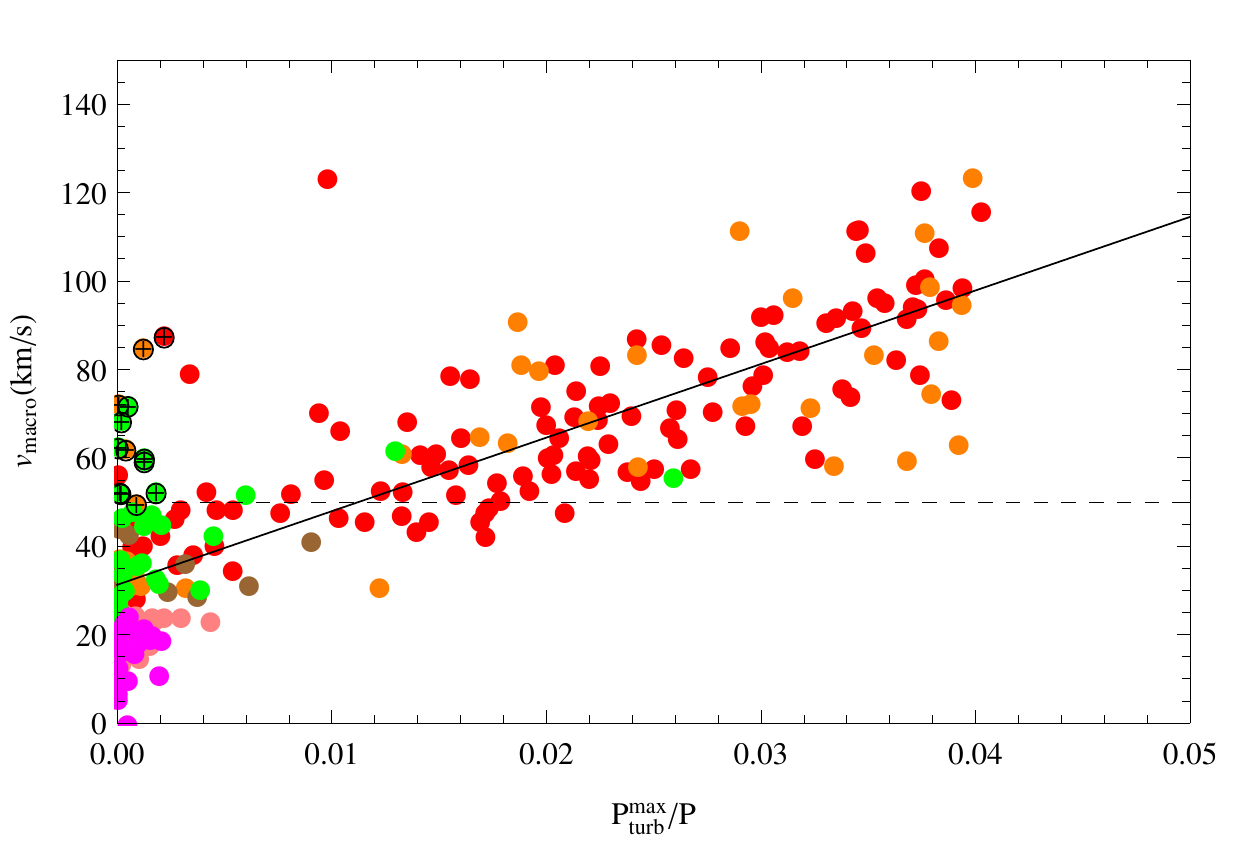}}
\caption{$v_{macro}$ as a function of the maximum fraction of turbulent-to-total pressure derived from a fitting of the tracks described in Sect.~3. The circles are color coded as in Fig.~\ref{fig:sHRmacro} with the crossed-black circled dots indicating the stars with $v_{macro}>50\,$km/s and log($\Ll)<3.3$. The dashed line indicates the 50~km/s level (see Fig.~\ref{fig:sHRmacro}).}
\label{fig:Pturbmacro}
\end{figure}

The agreement becomes even more striking looking at Fig.~\ref{fig:Pturbmacro}, where the observed macroturbulent velocities are directly plotted against the maximum fraction of turbulent pressure in our models. This plot reveals a clear correlation between macroturbulent velocity and turbulent pressure starting from $\Pt^{max}/P\approx0.005$, with a Spearman's rank correlation coefficient of 0.812.

\section{The connection to high-order pulsations}

Figure~\ref{fig:sHRmacro} revealed about ten stars with 
log($\Ll)<3.3$ for which our models predict very small
turbulent pressure contributions, but which unambiguously 
show macroturbulence at a level above 50\,km/s (see also Fig.~\ref{fig:Pturbmacro}).
Interestingly, we find that all these stars are located inside or very near to the region where stars are expected to be
pulsationally unstable to high-order g-modes (\citealt{2007Miglio,1999Pamyatnykh}, see also \citealt{2015SimonDiaz}), which is drawn as a gray band in Fig.~\ref{fig:sHRmacro}.
Indeed, \citet{2009Aerts} showed that the collective effect of high-order non-radial pulsations may produce a velocity field in the spectra of hot stars which resembles closely what is observed as macroturbulence in the stars discussed in Sect.\,4. 

Consequently, we interpret these stars as affected by a macroturbulent broadening that can be explained via a heat-driven pulsational origin. However, the homogeneity of the spectroscopic signatures of macroturbulence
over the whole effective temperature range in the luminous stars calls for a single dominant mechanism to produce it (Sim{\'o}n-D{\'{\i}}az et al. in prep.). This is not the case when considering only classical ($\kappa$-mechanism) instability domains, which do not cover the full region where most of the stars showing a macroturbulent velocity field are observed \citep[see ][Godart et al. in prep.]{2015SimonDiaz}. \citet{2013Shiode} and \citet{2015Aerts} consider gravity-waves originating in the convective core as the cause of macroturbulence. Whereas \citet{2013Shiode} based on their massive star models find that the surface velocity fluctuations do not exceed 10\,m/s even in their most optimistic case, \citet{2015Aerts}, based on 2-D non-linear simulations of convection-driven waves in a modified 3\Msun model concluded that it might explain the macroturbulence observed in O-stars. 
  
Within our scenario, the Reynolds stresses associated with turbulent pressure induce 
uncorrelated turbulent pressure fluctuations in the form
\begin{equation}
\delta P_{turb}\sim\rho v_c^2\,\,,
\label{turbperturb}
\end{equation}
\citep{1990Goldreich,2005Grigahcene,2013Lecoanet,2013Shiode}
where $\delta P_{turb}$ is the Lagrangian pressure perturbation associated with the convective motions.
Such stochastic fluctuations at the percent level can produce a strong local deviation from hydrostatic equilibrium, 
and will thus naturally excite high-order pulsations in the range of eigenmodes, which are closest to
the spectrum of the fluctuations. 

This would imply that also in the luminous stars with log($\Ll)>3.3$ 
in Fig.~\ref{fig:sHRmacro}, the macroturbulence may be signifying high-order pulsations, which are, however, excited by turbulent pressure fluctuations rather than through the $\kappa$-mechanism or by strange mode instability. If so, we can on one hand expect that 
linear pulsation analyses which include the Reynolds stress tensor, as e.g., in \citet{2004Dupret,2005Dupret} or \citet{2014Antoci}, may uncover
that stars in a large fraction of the red and orange coloured region in Fig.~\ref{fig:sHRmacro} 
are unstable to high-order g-mode pulsations. On the other hand, as the pressure fluctuations in these stars, as predicted by our simple analysis, can amount up to 5\% of the total pressure, it is conceivable that
in linear stability analyses, which require the growth of the instability from infinitesimal perturbations, 
an instability is not detected in all stellar models where high-order g-mode can be excited through 
finite amplitude pressure perturbations.    

\section{Conclusions}

We implemented the effect of the turbulent motion of convective eddies in a simple formalism
in the momentum and energy equations of our stellar evolution code. By comparing to previous
computations \citep{2011Brott}, we find that 
the turbulent pressure does not alter the stellar structure significantly. However, we find
maximum turbulent pressure contribution of up to 5\% and 30\% in our models for O-B supergiants
and cool red supergiants, respectively.
By comparing the maximum turbulent pressure contribution
in our models with spectroscopically derived macroturbulent velocities
for a large sample of Galactic OB stars \citep{2015SimonDiaz}, we find both quantities to be strongly correlated.

Several less luminous stars, in which the turbulent pressure is thought to be small,
show nevertheless high macroturbulent velocities. These are located close to or inside the region where linear pulsation analysis predicts high-order g-mode pulsations, arguing therefore for $\kappa$-mechanism pulsations, and not turbulent pressure fluctuations, as the origin of the macroturbulence phenomenon, in line with previous suggestions \citep{2009Aerts,2015SimonDiaz}.

We argue that the turbulent pressure fluctuations in hot luminous stars can excite such high-order pulsations, most likely non-radial g-modes, which may explain the occurrence of macroturbulence in stars which are found outside of the currently predicted pulsational instability domains. 
This view is in agreement with the indication that macroturbulence can be suppressed in strongly magnetic stars, given that such a field might effectively inhibit convective motions in the FeCZ \citep{2013Sundqvist}. 

At the moment, turbulent pressure fluctuations appear to
be the only mechanism which may excite high-order oscillations in luminous stars (log(L/L{$\odot$}$)>4.5$) in the wide
effective temperature regime for which strong macroturbulence is observed.

\begin{acknowledgements}
{\it LG is member of the International-Max-Planck-Research-School (IMPRS) for
Astronomy and Astrophysics at the Universities of Bonn and Cologne. LF acknowledges financial support from the Alexander-von-Humboldt foundation. SS-D acknowledges funding by the Spanish Ministry of Economy and Competitiveness (MINECO) under the grants AYA2010-21697-C05-04, AYA2012-39364-C02-01, and Severo Ochoa SEV-2011-0187. Based on observations made with the Nordic Optical Telescope, operated by NOTSA, and the Mercator Telescope, operated by the Flemish Community, both at the Observatorio del Roque de los Muchachos (La Palma, Spain) of the Instituto de Astrof\'isica de Canarias.
 Moreover, LG and LF thank A.~Miglio for useful discussions. }
\end{acknowledgements}


\end{document}